\documentstyle[eqsecnum,aps,epsf,twocolumn]{revtex}

\begin{document}

\preprint{DUKE-TH-98-169}

\title{Gluon Field Dynamics in Ultra-Relativistic Heavy-Ion Collisions:
Time evolution on a Gauge Lattice in 3+1 Dimensions }

\author{W. P\"oschl and B. M\"uller   \\
Department of Physics, Duke University, Durham, NC 27708-0305, USA }
\vspace{15mm}
\date{\today}
\maketitle
\vspace{15mm}

\maketitle

\begin{abstract}
We describe the dynamics of gluons and quarks in a relativistic nuclear
collision, within the framework of classical mean-field transport theory,
by the coupled equations for the classical Yang-Mills field and a collection
of colored point particles.  The particles represent the valence quarks in
the colliding nuclei. We explore the time evolution of the gauge field
in a numerical simulation of the collision of two Lorentz-boosted ``nuclei'' 
on a long three-dimensional gauge lattice. We report first results on soft 
gluon scattering and coherent gluon radiation.
\end{abstract}
\bigskip\bigskip


Experiments with relativistic heavy ions at center-of-mass energies reaching
100 GeV/u will soon search for a new phase of nuclear matter, the quark-gluon 
plasma \cite{Harris.96,Smilga.97}. One of the theoretical challenges in this 
context is the development of a description on the basis of quantum 
chromodynamics (QCD)
of the processes that may lead to the formation of locally equilibrated
superdense matter in these nuclear reactions.

The possibility of a description of inelastic gluon processes by means
of the nonlinear interactions of classical color fields has been proposed
some time ago \cite{Ehtamo.83,Kovner.95}. Recently, this scenario was
examined in studies of the collision of two transverse polarized Yang-Mills 
field wave packets on a one-dimensional gauge lattice \cite{Hu.95}. These
calculations showed that the interaction between localized classical 
gauge fields can lead to the excitation of long wavelength modes in a way 
that is reminiscent of the formation of a dense gluon plasma.

Here we address the question to which extent soft gluon scattering
in ultra-relativistic nuclear collisions contributes to the excitation
of gluonic modes with non-vanishing transverse momentum and, therefore,
to the conversion of longitudinal kinetic energy into transverse energy. 
The colliding gluon fields are generated by classical color sources, 
representing the fast moving nuclear valence quarks. The inclusion of 
particles also permits a comparison with perturbative calculations of 
gluon radiation from colliding quarks \cite{Kovchegov.97,Matinyan.97}.

A set of equations describing the evolution of the phase space
distribution of quarks and gluons in the presence of a mean color field,
but in the absence of collisions, was proposed by 
Heinz \cite{Heinz.83,Elze.89}. 
This non-Abelian generalization of the 
Vlasov equation can be considered as the continuum version of the dynamics
of an ensemble of classical point particles endowed with color charge
and interacting with a mean color field. Denoting the space-time positions,
momenta, and color charges of the particles by $x_i^{\mu}$, $p_i^{\mu}$ and 
$q_i^a$, respectively, where $i=1,\ldots,N$ is the particle index, the
equations \cite{Wong.70} for this dynamical system read:
\begin{eqnarray}
\label{Eq.1}
m{{dx_i^{\mu}}\over{d\tau}} &=& p_i^{\mu} \\
\label{Eq.2}
m{{dp_i^{\mu}}\over{d\tau}} &=& gq_i^aF^{\mu\nu}_a p_{i,\nu} \\
\label{Eq.3}
m{{dq_i^{a}}\over{d\tau}}   &=& - g f^{abc} q_i^b p_i^{\mu} A_{\mu}^c .
\end{eqnarray}
Here $g$ is the gauge coupling constant, $f^{abc}$ are the structure 
constants of the gauge group (here taken as SU(2)), and $F_a^{\mu\nu}$ 
denotes the strength of the mean color field $A_{\mu}^c$. 
The moving particles generate a color current ${\cal J^{\mu}}=J^{\mu}_a
\tau^a/2$ which forms the source term of the inhomogeneous Yang-Mills 
equations 
\begin{eqnarray}
\label{Eq.5}
D_{\mu}{\cal F}^{\mu\nu}(x)\, =\, g\,{\cal J}^{\nu}(x)\,=\,
g\,\sum_i {\cal Q}_i(t) {{p_i^\nu}\over{m}} \delta(\vec x-\vec x_i(t)) ,
\end{eqnarray}
for the mean color field. 
These equations have recently been used in the weak-coupling limit 
($g\ll 1$) to simulate the effects of hard
thermal loops \cite{Pisarski.89} on the dynamics of soft modes of a
non-Abelian gauge field at finite temperature \cite{Hu.97,Moore.98}.
In this case, the colored particles describe the gauge field modes with
thermal momenta, and the mean field describes the coherent motion of those
gauge field modes which have a wave number $k$ much smaller than the
temperature $T$ and are highly occupied. 

Here we use Eq. (\ref{Eq.1}--\ref{Eq.5}) to describe the 
interactions among the soft glue field components of two colliding heavy 
nuclei. Transverse modes are included by using a 3-dimensional spatial
lattice.  The short-distance lattice cut-off $a$ separates the
regime in transverse momentum where the dynamics of gluons is perturbative
(large $k_{\rm T}$) from that where perturbation theory fails (small 
$k_{\rm T}$) and which is of interest to us here.
The interaction with the mean color field allows for an exchange of an 
arbitrary number of soft gluons. In combination with a parton cascade,
the screening of the soft components of the gauge field by perturbative partons
\cite{Biro.92,Eskola.96} is taken into account naturally by the nonlinear
nature of the coupled Eq. (\ref{Eq.1}--\ref{Eq.5}). This will be
discussed in a separate publication \cite{BMP.98}.

We represent the valence quarks of the 
two colliding nuclei as point particles moving in the space-time continuum,
and interacting with a classical gauge field defined on a spatial lattice
in continuous time. Here we neglect the collision integrals describing 
hard interactions between the particles. In the spirit of the statistical
nature of the transport theory, we split each quark into a number $n_q$ of 
test particles, each of which carries the fraction $q_0=Q_0/n_q$ of the 
quark color charge $Q_0$. For the gauge group SU(2) adopted here, 
each nucleon is represented by two quarks, initially carrying 
opposite color charge.  A lattice version of the continuum equations 
(\ref{Eq.1}--\ref{Eq.5}) is constructed 
\cite{Hu.97} by expressing the field amplitudes as elements of the 
corresponding Lie algebra, i.e. ${\cal A}_{\mu},{\cal F}_{\mu\nu},
{\cal E}_k,{\cal B}_k\,\in$ LSU(2).  As in the Kogut-Susskind model 
of lattice gauge theory \cite{Kogut.75} we choose the temporal gauge 
${\cal A}_0 = 0$ and define the following variables. 
\begin{eqnarray}
\label{Eq.6}
{\cal U}_{x,l} 
&=& \exp(-iga_l{\cal A}_l(x)) \,\,=\,\, {\cal U}^{\dagger}_{x+l,-l} \\
\label{Eq.7}
{\cal U}_{x,kl} 
&=& {\cal U}_{x,k}\, {\cal U}_{x+k,l}\, {\cal U}_{x+k+l,-k}\, 
                  {\cal U}_{x+l,-l} 
\end{eqnarray}
Consequently, we have
\begin{eqnarray}
\label{Eq.8}
{\cal E}_{x,j} = { {-i}\over{ga_j}}\,\dot {\cal U}_{x,j}
                   {\cal U}^{\dagger}_{x,j}\, ,    \quad
{\cal B}_{x,j} = { {i\,\epsilon_{jkl}}\over{4ga_ka_l}}\,
\bigl({\cal U}_{x,kl} - 
{\cal U}^{\dagger}_{x,kl} \bigr)
\end{eqnarray}
for the electric and magnetic fields, respectively. The lattice constant
in the spatial direction $j$ is denoted by $a_j$.  As one can see from 
(\ref{Eq.6}), the gauge field is expressed in terms of the link variables 
${\cal U}_{x,l}\,\epsilon$ SU(2), which represent the parallel transport 
of a field amplitude from a site $x$ to a neighboring site $(x+l)$ in 
the direction $l$.  We choose ${\cal U}_{x,i}$ and ${\cal E}_{x,i}$ as the 
basic dynamic field variables and solve the following equations of motion
\begin{eqnarray}
\label{Eq.10}
\dot{\cal U}_{x,k}(t) = i\,g\,a_k\,{\cal E}_{x,k}(t)\,{\cal U}_{x,k}(t) 
\qquad\qquad\qquad\qquad\qquad \\
\label{Eq.11}
\dot{\cal E}_{x,k}(t) = -\,{\cal J}_{x,k}(t) \, + \,
{i\over{2ga_1a_2a_3}}
\sum\limits^3_{l=1}
\Bigl\{{\cal U}_{x,kl}(t) -
{\cal U}^{\dagger}_{x,kl}(t) 
\,\Bigr.  \\
-\,  \Bigl. {\cal U}^{\dagger}_{x-l,l}(t)\,
\Bigl( {\cal U}_{x-l,kl}(t) -
{\cal U}^{\dagger}_{x-l,kl}(t)\Bigr)\,
{\cal U}_{x-l,l}(t)\, \Bigr\}.
\nonumber 
\end{eqnarray}

Subsequently, we present results of a calculation which corresponds
to a central collision of two Pb nuclei. 
We assume that 16 nucleons in a row collide on the collision
axis (3-axis). In accordance with the spatial extension of a nucleon
we choose a lattice extension of 1.2~fm into both transverse 
directions. The lattice spacing is taken $a_j = 0.3$ fm in
each direction, thus $4^2\times 40$ lattice points cover the volume 
of 8 nucleons in one row. The coverage of two complete nuclei
requires much larger lattices and remains a challenge for more
extended calculations in the future. The lattice is closed to a 
3-torus and a dual lattice is superimposed on the original lattice in such 
a way that the lattice points are located in the centers of the cells of 
the dual lattice. The dual lattice cells are used to associate particles
with lattice sites and thus to define the source terms in (\ref{Eq.5}).

To generate the initial configurations of the nuclei we randomly distribute 
color charged massless particles over the volume such that each lattice cell
is occupied by an even number $n_b$ of particles. The total initial
charge is zero in each box corresponding to a neutral charge distribution.
Momenta with opposite but random directions and Boltzmann distributed 
absolute values are assigned to each pair of particles.  The equilibrium 
configuration of a single nucleus in its rest frame is obtained
through the evolution of (\ref{Eq.1}--\ref{Eq.3}, \ref{Eq.10},
\ref{Eq.11}) over a long period of time starting from the  
initial fields ${\cal E}_{x,k}(t=0) = 0$, ${\cal U}_{x,k}(t=0) = {\bf 1}_2$. 
As shown in Table 1, we find ratios between field energy $W_{\rm f}$ 
and total energy $W_{\rm tot}$ which descend smoothly from 
$W_{\rm f}/W_{\rm tot} = 0.45$ to $W_{\rm f}/W_{\rm tot} = 0.14$ 
for particle densities $n_b = 2$ to $n_b = 10$. In Table 1 
we also demonstrate that the ratio does not depend on $a$, $g$, and $Q_0$.
The ratio is also independent of the time step (if $\Delta t \le a/10 $). 
For the large final times $t_n\approx 10^4$, the value of 
$W_{\rm f}/W_{\rm tot}$ has saturated, except for very small couplings, 
such as $g=0.3$, where the equilibrium is reached only after longer times. 

The simulation of a collision requires a Lorentz boost of each nucleus
into the center of velocity frame of both nuclei. In the example presented
below, the kinetic energy is 100 GeV/u for which $\gamma = 106.5$.
Both nuclei are mapped
into their initial positions on a large Lorentz-contracted lattice right
after the boost of particle coordinates $x_i, p_i$ and field amplitudes.
We have investigated two ways to perform the boost. In one method we
thermalise the nuclei first and apply the Lorentz transformation 
afterwards. This method is complicated and requires a continuation of 
the time-evolution throughout the boost.  Furthermore, the lattice 
dispersion relation is not Lorentz invariant and therefore 
it does not lead to a stable translation of the nuclei
on the lattice after the boost. 

More successful is a boost of the particle coordinates $x_i, p_i$ 
at time $t=0$ for both nuclei, wherafter the fields are generated 
from the initial conditions ${\cal E}_{x,k}(t=0) = 0$ and 
${\cal U}_{x,k}(t=0) = {\bf 1}_2$. This method leads to 
quasi-stable configurations propagating over distances which
extend over several thousand lattice points in the longitudinal direction. 
However, it requires large initial distances between the
nuclei because an equilibrium between particles and fields has
to be established before the collision. For nucleons with the
correct mass ($m_n = 939$ MeV) about $50\%$ of the total energy are
carried by glue fields.
For the above chosen lattice constants each nucleon is covered by
80 lattice cells and thus, according to the results in Table 1, each quark 
must be split into $n_q=160$ particles to obtain a ratio close to
$W_{\rm f}/W_{\rm tot}= 0.5$ at equilibrium. In order to
reproduce the correct nucleon mass the particle momenta 
have to be smaller by a factor $1/n_q$ on average in comparison to 
the momenta of the quarks. The field energy associated with
a single link can then occasionally exceed the kinetic energy of 
a particle resulting in the emission of particles due to backwards
scattering from the link. This corresponds to the (unphysical) 
emission of nucleons from the moving nuclei, which we want to avoid.

As long as we are only interested in the dynamics of the glue fields 
we may increase the momenta of the particles such that the particle
distributions of the nuclei remain stable throughout the whole collision.
Therefore, in order to prepare a stable initial state for the collision, 
we start (before the Lorentz boost) with an initial condition 
in which the momenta of the particles are generated at 
a fictitious temperature of $T_p=100$ GeV.
The directions of the 
momenta are randomly distributed into the forward hemisphere such
that pairs of particles have opposite transverse momenta. 
We found that the final results are quite independent of
the initial condition.  These initial conditions, however, contain 
much higher total energy $W'_{\rm tot}$ than $W_{\rm tot}$ and are 
unphysical.  The fact that the temperature of the
gluon field $T\ll T_p$ is limited by 
$T_{max}= 1.3$ GeV due to the lattice cutoff prevents
the field energies from becoming too large. 
Furthermore, we increase the value for $n_b$ to keep the ratio 
$W_{\rm f}/W_{\rm tot}$ small while $W_{\rm tot} \ll W'_{\rm tot}$ .  
For $n_b=6 $ choosen, an initial separation $d_0$ of more
than 12000 lattice spacings is necessary to approach a ratio of
about $W_f/W_{\rm tot}=0.5$. $d_0$ is Lorentz contracted to $33.8$ fm 
in the center of velocity frame. According to the above fixed
extensions into transverse directions and size of the nuclei
we use a lattice that comprises a total of $4^2\times 12100$ points.
While the nuclei translate on the lattice, we adiabatically increase the 
coupling constant with a rate of $\Delta g/\Delta t \le (5000a)^{-1}$. 
After $g$ has reached its final value ($g=2.0$), we continue the 
time evolution until $W_{\rm f}/W_{\rm tot}\approx 0.5$. 
\vskip 0.2cm
%
%
%
%
\begin{figure}[H]
\centerline{
\epsfysize=8cm \epsfxsize=8cm                  
\epsffile{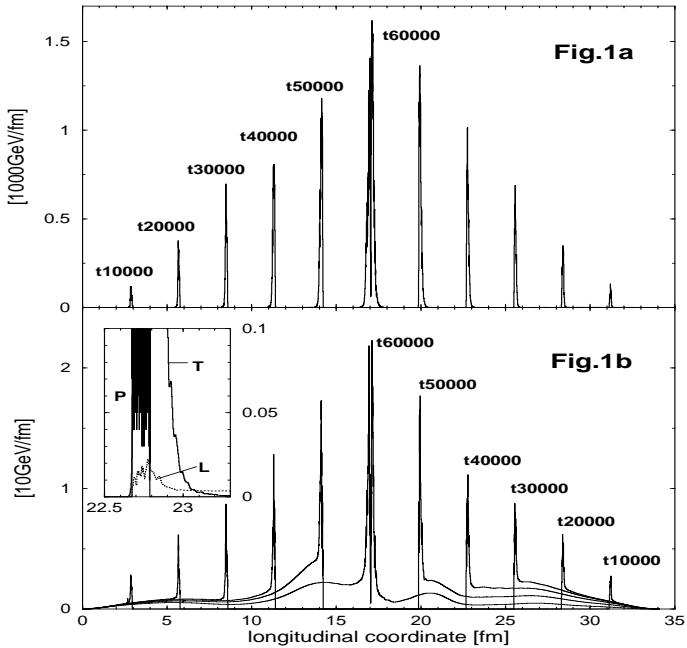} 
}
\vskip 0.3cm
\caption{Transverse and longitudinal $E$-field energy densities at
various times before the collision.}
\end{figure}
\noindent
In the following, we call the direction $j=3$ which is parallel to
the collision axis the ``longitudinal'' direction and $E$-field
amplitudes associated with longitudinal links are called the
``longitudinal fields'' $E_{\rm l}$. 
The fields $E_{\rm t}$ in transverse directions are called ``transverse''.
Figures 1a,b display the field energy densities 
$w_{\rm t}^{(E)}(z) = Tr({\cal E}_1 {\cal E}_1) + Tr({\cal E}_2 {\cal E}_2)$,
(see Fig. 1a) and  
$w_{\rm l}^{(E)}(z) = Tr({\cal E}_3 {\cal E}_3)$ (see Fig. 1b) for both,
transverse and longitudinal  
compontents of the $E$-field every 10000 time steps. 
A time-step width of 0.03 fm has been used. 
We verified, that the corresponding $B$-field energy densities
$w^{(B)}(z)$ are identical with $w^{(E)}_{\rm t}(z)$. After
1000 time steps (not shown in the figure)
the transverse $E$-field energy $W^{(E)}_{\rm t}$ is by about a 
factor $\gamma/2$ larger than the longitudinal $E$-field 
energy $W^{(E)}_{\rm l}$.  This ratio increases to $\gamma$ at 50000 
time steps and remains constant afterwards. 

As Fig. 1a shows, two configurations representing two nuclei 
propagate remarkably stably (nucleus 1 from left to right and
nucleus 2 from right to left) over long distances on the lattice. 
In the upper left corner of Fig. 1b
at time step 40000, the particle density (P), 
which is defined as the number of particles per bin of width $a/10$
in longitudinal direction, is superimposed on the transverse (T) 
and longitudinal (L) field energy densities for comparison. 
The particle density defines the extension of the nucleus in
the longitudinal direction. The mismatch between particle density 
and field energy density decreases when the density 
of lattice points covering the nucleus in the longitudinal direction is
increased. This requires a shift of the cutoff (here at $k_c=2.0$ GeV 
for $a=0.3\,{\rm fm}$) to higher momenta.  

The tails of the field energy densities 
result not only from lattice dispersion effects. 
The lattice is Lorentz contracted in the longitudinal direction and 
particles have large longitudinal momentum components. Therefore, 
particles primarily transfer energy into longitudinal links.
The nonlinearity of the Yang Mills
equations provides a mechanism transferring energy from 
longitudinal into transverse degrees of freedom. Field amplitudes
which are left behind on the longitudinal links are cancelled by 
following particles of rotated color charge. 
Fig. 1b shows that this cancellation does not work perfectly, and 
small, but long tails remain behind the nuclei. 
The amplitude of these tails decreases for increasing $n_b$ and $d_0$.
Another advantage of choosing $n_b$ large is to avoid artificial 
excitation of high frequency modes caused by an interplay of 
lattice discretization and point-like charges.

Fig. 2a and Fig. 2b display $w^{(E)}_{\rm t}(z)$ and $w^{(E)}_{\rm l}(z)$ 
at 4000, 8000, 12000 and 16000 time steps after the collision. 
The particle densities are superimposed to indicate the position 
and original extension of the nuclei. As compared to Fig. 1b (upper
left window) the width of the distributions $w^{(E)}_{\rm t}(z)$ 
moving with the particles is considerably increased. At the 
begining of the collision we observe a kink in $W_{\rm f}(t)$, which
increases remarkably in the collision. As we see from Fig. 2a, a 
considerable fraction of the transverse field energy is deposited around 
the center of collision between the receding nuclei.
At time step 76000,
about $30.1\,{\rm GeV}$ of the total transverse electric field energy    
$W^{(E)}_{\rm t}= 490.7\,{\rm GeV}$ (at time step 60000) 
are left in the spatial region 
between $13.9\,{\rm fm}$ and $19.9\,{\rm fm}$. Consequently, the
energy density in the overlap region with the particles 
is reduced allowing an enhanced transfer of energy from the
particles into the fields. This results in the kink in $W_{\rm f}(t)$.

Fig. 2b displays the corresponding longitudinal energy densities.
The calculation shows that within
short times ($\Delta t = 0.5\,{\rm fm}$) 
after the collision about $26\,\%$ ($7.25\,{\rm GeV}$) of the total field
energy ($28.03\,{\rm GeV}$) left
in the region between $16.4\,{\rm fm}$ and $17.4\,{\rm fm}$
are carried by longitudinal components of the $E$-field. The ratio 
$w^{(E)}_{\rm l}/w^{(E)}_{\rm t}$ in this region is 0.7 which is
large compared to the above mentioned factor $1/\gamma$. Since
the color charge and color current density is zero, the linearized
equations (\ref{Eq.5}) are homogeneous in this region and
allow only for solutions with non-zero amplitudes into
transverse directions in relation to the direction of their
energy flow. Figure 2b displays the fraction
with momenta pointing into transverse direction relative to
the collison axis. A calculation with the right-hand side of 
(\ref{Eq.2}) set to zero yields practically the same results,
which finds its explanation in a transfer of energy from the propagating 
transverse fields to the longitudinal fields due to the non-linear 
coupling between $E_{\rm t}$ and $E_{\rm l}$ in (\ref{Eq.5}). When the
two nuclei overlap during the collision, the fields are 
superposed. As a result of the nonlinear terms in (\ref{Eq.5}),
which act as source terms for the longitudinal fields, the  
amplitudes $E_{\rm l}$ start to grow rapidly.
This mechanism describes the scattering of soft gluons.
\vskip 0.3cm
%
%
%
%
%
\begin{figure}[H]\centerline{
\epsfysize=8cm \epsfxsize=7cm                  
\epsffile{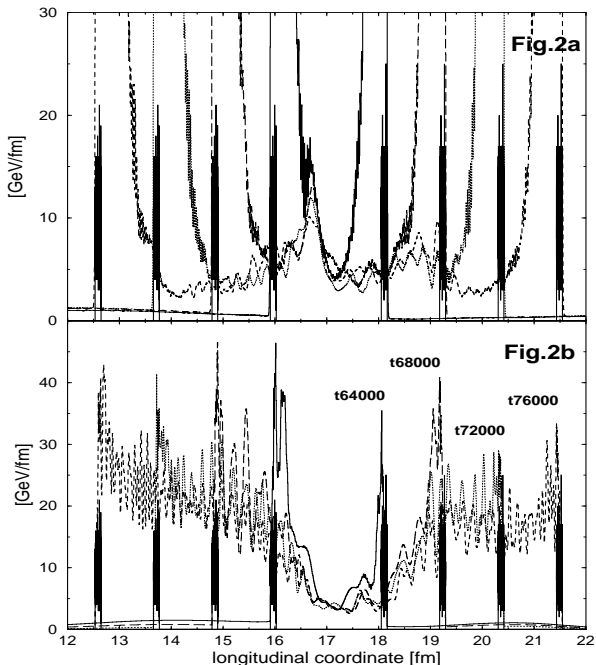} 
}
\vskip 0.3cm
\caption{Transverse and longitudinal $E$-field energy densities
after the collision.}
\end{figure}
\noindent
The distribution $w_{\rm l}^{(E)}(z)$ 
increases sharply for collision times larger $1.0\,{\rm fm}$.
When the two nuclei pass through each other, the particles experience 
the combined fields of both nuclei. The corresponding changes of the
field amplitudes enter into the r.h.s. of (\ref{Eq.3})
and modify the orientation of the color charge vectors $\vec q_i^a(t)$
during the rather short overlap time of $0.11\,{\rm fm}$.
This induces net color charge currents 
resulting in radiation of gluons. Since the longitudinal momenta of
the particles are large compared with their transverse components,
color charges move essentially parallel to longitudinal links
and induce electric fields on these links. Before the collision,
the charge of the following particles was polarized such
that these fields were cancelled. After the collison 
this is no longer true. A continuation of
the time evolution has shown that the longituinal energy
density of the fields remains as large as in Fig. 2b and
decreases slowly after time step 100000 ($28.17\,{\rm fm}$).
These longitudinal fields possess only transverse momenta
and result in gluon radiation into transverse directions. 

The energy distribution in Fig. 2a and in particular the hump 
at the center of collision
result from non-linear properties of Eq. (\ref{Eq.5}) and exhibit 
the gluon-gluon interaction. 
We repeated the calculation for smaller initial distances $d_0$.
It turned out that there is a threshold between $7000a$ and
$9000a$ below of which the hump doesn't appear and 
$w_{\rm t}^{(E)}(z)<w_{\rm l}^{(E)}(z)$ between the receding nuclei. Further, 
$w_{\rm t}^{(E)}(z)$ is almost constant as a function of z in this case.
As can be seen from Fig. 2a and 2b, above this threshold, we find 
$w_{\rm t}^{(E)}(z)>w_{\rm l}^{(E)}(z)$ in a small region around the 
center of collision.
The slow decay of the hump indicates that essentially
low frequent modes are excited in the collision of the fields.
The transverse fields have momenta into longitudinal directions
and thus describe longitudinal gluon radiation.
A comparison of the transverse field energy deposit with
the longitudinal field energy deposit for large times ($t/a=76000$)
shows that the radiation of soft gluons into longitudinal
directions amounts to about $30\%$ of the intensity into
transverse directions.   

The authors are thank S.A. Bass and S.G. Matinyan for discussions.
This work was supported by the U.S. Department of Energy under
Grant No. DE-FG02-96ER40495.



%
%
%
\begin{table}
\begin{center}
\begin{tabular}{|c|c|c|c|c|c|}

$a$ [fm] & $g$ & $Q_0$ & $n_b$ & $t_n$ [a/10] & $W_{\rm f}/W_{\rm tot}$ \\
\hline
  1.0        & 1.0 & 2.0  &  2     &$1\cdot 10^4$&0.45\\
  1.0        & 1.0 & 2.0  &  4     &$1\cdot 10^4$&0.30\\
  1.0        & 1.0 & 2.0  &  6     &$1\cdot 10^4$&0.26\\
  1.0        & 1.0 & 2.0  & 10     &$1\cdot 10^4$&0.14\\ 
\hline
  0.1        & 1.0 & 2.0  &  2     &$1\cdot 10^4$&0.45\\
  0.1        & 1.0 & 2.0  &  4     &$1\cdot 10^4$&0.30\\ 
\hline
  0.3        & 1.0 & 2.0  &  2     &$1\cdot 10^4$&0.45\\
  0.3        & 2.0 & 2.0  &  2     &$1\cdot 10^4$&0.46\\
  0.3        & 3.0 & 2.0  &  2     &$1\cdot 10^4$&0.46\\ 
\hline
  0.3        & 0.5 & 2.0  &  2     &$1\cdot 10^4$&0.37\\
  0.3        & 0.5 & 2.0  &  2     &$2\cdot 10^4$&0.44\\ 
\hline
  0.3        & 0.3 & 2.0  &  2     &$1\cdot 10^4$&0.19\\
  0.3        & 0.3 & 2.0  &  2     &$3\cdot 10^4$&0.36\\
  0.3        & 0.3 & 2.0  &  2     &$6\cdot 10^4$&0.45\\ 
\hline
  0.3        & 2.0 & 1.0  &  2     &$1\cdot 10^4$&0.46\\
  0.3        & 2.0 & 0.5  &  2     &$1\cdot 10^4$&0.46\\
  0.3        & 2.0 & 0.3  &  2     &$1\cdot 10^4$&0.46

\end{tabular}
\end{center}
\caption{The ratio $W_{\rm f}/W_{\rm tot}$ is listed in dependence
on the parameters $a$, $g$, $Q_0$, $n_b$ for large times $t_n$
when the configuration has evolved close to equilibrium.}
\end{table}

\end{document}